\documentclass[%
 reprint,
 superscriptaddress,
 amsmath,amssymb,
prx,
longbibliography
]{revtex4-2}

\usepackage{graphicx}
\usepackage{dcolumn}
\usepackage{bm}

\usepackage{natbib}
\usepackage[colorlinks,allcolors=blue]{hyperref}
\usepackage[dvipsnames]{xcolor}
\usepackage{float}
\usepackage{physics}

\graphicspath{{figures/}}

\begin{document}


\title{Robust preparation of ground state phases under noisy imaginary time evolution}

\author{Aleksei Khindanov}
\email{akhin@ameslab.gov}
\affiliation{Ames National Laboratory, U.S. Department of Energy, Ames, Iowa 50011, USA}
 
\author{Yongxin Yao}
\affiliation{Ames National Laboratory, U.S. Department of Energy, Ames, Iowa 50011, USA}
\affiliation{Department of Physics and Astronomy, Iowa State University, Ames, Iowa 50011, USA}

\author{Thomas Iadecola}
\email{iadecola@iastate.edu}
\affiliation{Ames National Laboratory, U.S. Department of Energy, Ames, Iowa 50011, USA}
\affiliation{Department of Physics and Astronomy, Iowa State University, Ames, Iowa 50011, USA}

\date{\today}

\begin{abstract}
Non-unitary state preparation protocols such as imaginary time evolution (ITE) offer substantial advantages relative to unitary ones, including the ability to prepare certain long-range correlated states more efficiently. 
Here, we ask whether such protocols are also robust to noise arising due to coupling to the environment.
We consider a non-unitary ITE ``circuit" subjected to a variety of noise models and investigate whether the resulting steady state remains in the same phase as the target state of the ITE at finite noise strength.
Taking the one-dimensional quantum Ising model as a concrete example, we find that the ground state order and associated phase transition persist in the presence of noise, provided the noise does not explicitly break the symmetry that protects the phase transition.
That is, the noise must possess the protecting symmetry in a weak (or average) form.
Our analysis is facilitated by a mapping to an effective Hamiltonian picture in a doubled Hilbert space.
We discuss possible implications of these findings for quantum simulation on noisy quantum hardware.
\end{abstract}

\maketitle

\section{\label{sec:Intro}Introduction}

High-caliber ground state preparation is a necessary step for many quantum chemistry and quantum many-body physics algorithms, most notably quantum phase estimation \cite{pea_kitaev,nielsen2002quantum,bauer2020quantum,dalzell2023quantum_arxiv}.
While various approaches to the ground state preparation have been proposed, such as adiabatic \cite{farhi2000quantum_arxiv,asp,Albash2018_AQC} and variational \cite{peruzzoVariationalEigenvalueSolver2014,vqe_theory,alan_ucc2018} state preparation, many of them suffer from limitations, particularly when being executed on near-term devices where resources for quantum computation are scant.
These limitations can include a small energy gap along the adiabatic path, the dependence on a variational ansatz or an ill-behaved classical optimization subroutine.

Another conceptual approach to ground state preparation, which can potentially be more suitable for near-term implementation, is based on imaginary time evolution (ITE) \cite{bauer2020quantum,qite_chan20,sun2020quantum,QITE_h2,qite_nla,Kamakari2022_qite_open,qite_qft,hejazi2024adiabatic_arxiv,VQITE,Jones2019_vqite,Endo20variational,theory_vqs,AVQITE,smqite,benedetti2020hardware,Chen2024_avqite_open,Liu2021_pite,lin2021real,Kosugi2022_PITE,mao2023}.
Being a non-unitary operation, the ITE cannot be implemented using only unitary gates; nevertheless, various algorithms have been designed that \textit{effectively} realize an approximate ITE process on a quantum device.
These algorithms either (1) perform local tomography to determine a unitary such that the unitarily-evolved state approximates the ITE-evolved state \cite{qite_chan20,sun2020quantum,QITE_h2,qite_nla,Kamakari2022_qite_open,qite_qft} (this is known as quantum imaginary time evolution, or QITE), (2) classically construct an adiabatic protocol that emulates the ITE \cite{hejazi2024adiabatic_arxiv} (this is known as adiabatic quantum imaginary time evolution, or A-QITE), (3) take advantage of the variational principle \cite{VQITE,Jones2019_vqite,Endo20variational,theory_vqs,AVQITE,smqite,benedetti2020hardware,Chen2024_avqite_open} (this is known as variational quantum imaginary time evolution, or VQITE), (4) utilize ancilla(e) and measurements \cite{Liu2021_pite,lin2021real,Kosugi2022_PITE} (this is known as probabilistic imaginary time evolution, or PITE), or (5) ancilla(e), measurements, and feedback \cite{mao2023}.
These and other sophisticated protocols become increasingly realizable on present-day and near-term devices due to rapid hardware advances, such as the ever more efficient implementation of mid-circuit measurements and the adoption of dynamical circuits \cite{Koh2023,iqbal2023_toric_code_quantinuum_arxiv,fossfeig2023_advantage_adaptive_circuits_quantinuum_arxiv}.

\begin{figure}
	\includegraphics[width=0.65\columnwidth]{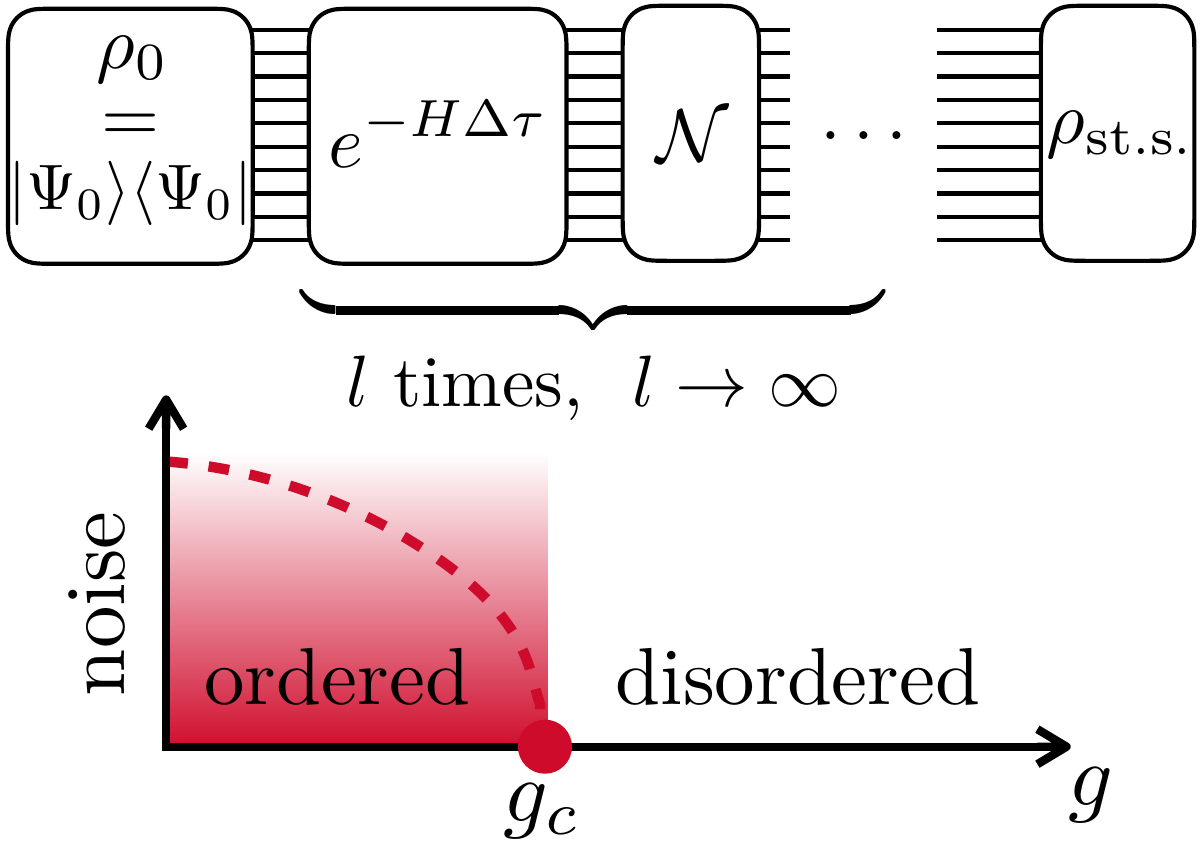}
	\caption{(Top) Schematic of the model studied in this paper. An initial pure state is fed into a ``circuit" consisting of black-box ITE (represented in a Trotterized form) followed by a noise channel $\mathcal N$. The object of interest is the steady state density matrix obtained upon iterating this evolution. (Bottom) When the Hamiltonian $H$ has a ground-state phase transition at a parameter $g=g_c$, this becomes a dynamical phase transition in the steady state of the ITE circuit in the limit of zero noise. For certain noise channels, it is possible for this phase transition to persist in the presence of noise, giving rise to a two-dimensional phase diagram.}
	\label{fig:schematic}
\end{figure}

So far the design of the ITE algorithms has primarily focused on the efficiency of their implementation on near-term machines---the proposed protocols attempt to minimize both the number of qubits required for the algorithm and the number of circuit layers, including measurements. 
However, another factor that has so far been mostly overlooked in studies of ITE algorithms is their stability to quantum noise arising due to coupling of the system to the environment.
In the absence of quantum error correction, the effects of noise are ubiquitous in near-term devices where it generally tends to damage quantum information and corrupt the performance of quantum algorithms \cite{nielsen2002quantum,preskill2015lecture}.

In this work we ask whether ITE-based ground state preparation is robust to various types of noise commonly present in experimental setups.
To keep our considerations as general and as simple as possible, we refrain from directly modeling a particular algorithm implementing the ITE, and instead consider a black box realizing a Trotterized version of ITE~\footnote{This Trotterized approach to ITE is also implemented directly in a variety of classical algorithms for quantum many-body physics, including time-evolving block decimation~\cite{Vidal03,Vidal04} and various quantum Monte Carlo approaches~\cite{Handscomb62,Blankenbecler83,Sandvik91,beach2019making}}.
A ``circuit" constructed in this way can be thought of as a toy model of an ITE protocol, which we further subject to noise (see Fig.~\ref{fig:schematic}).
Within this model the ITE effectively realizes an approximate (due to the Trotter error) weak projection onto a target ground state, which may belong to a nontrivial phase, whereas noise tends to trivialize the state it acts upon.
Here we analyze this competition between ITE and noise by investigating whether the mixed state resulting from the noisy ITE process belongs to the same nontrivial \textit{phase} as the targeted pure ground state.

As a concrete example, we focus on the one-dimensional transverse-field Ising model whose ground-state quantum phase transition is a prototypical example of long-range order and spontaneous symmetry breaking (SSB). 
We investigate whether this order, and the quantum critical point from which it emerges, persists in the mixed steady state of the ITE at finite noise strength. 
To this end, in the limit of infinitesimal imaginary Trotter step we take advantage of the Choi-Jamio\l kowski isomorphism \cite{Jamiolkowski1972,Choi1975} and map the noisy ITE of interest to a noiseless ITE under an effective Hamiltonian in a doubled Hilbert space.
This allows us to extract the steady state properties of the noisy ITE from the ground state of the effective Hamiltonian, which can be efficiently calculated using the density matrix renormalization group (DMRG)~\cite{White1992dmrg,White1993dmrg,Schollwock2005_dmrgRMP}.
To further corroborate our findings and investigate the effect of imaginary-time Trotter error, we also perform direct simulations of our noisy ITE ``circuit" for comparatively small system sizes.
We find that the ordered phase is stable to sufficiently weak noise as long as the noise does not explicitly break the weak version \cite{Buca_2012_strong_weak_sym, deGroot2022symmetryprotected} of the symmetry that is spontaneously broken at the ground-state quantum critical point. 

The rest of this paper is organized as follows.
In Sec.~\ref{sec:Model} we introduce the model and review the basics of the doubled Hilbert space formalism.
We present our main results in Sec.~\ref{sec:Results}, focusing on several illustrative examples of both Pauli (Sec.~\ref{sec:Pauli}) and non-Pauli (Sec.~\ref{sec:Amplitude_damping}) noise channels.
We conclude in Sec.~\ref{sec:Conclusion}, where we touch on possible implications of our results for ITE quantum algorithms implemented on noisy hardware.

\section{\label{sec:Model}Model}

We consider a Trotterized evolution along the imaginary time axis as a black box model for an ITE protocol.
Assuming that the target Hamiltonian can be decomposed into a sum of local terms as $H=\sum_m h_m$, the mixed-state density matrix $\rho$ after each Trotter step becomes
\begin{equation}
	\mathcal{I}_m[\rho]=\frac{e^{-\Delta\tau h_m}\rho e^{-\Delta\tau h_m}}{\Tr[e^{-2\Delta\tau h_m}\rho]},
 \label{eq:qite_trotter_m}    
\end{equation}
where $\Delta\tau$ is the size of the Trotter step. 
For small $\Delta\tau$ the composition of superoperators $\mathcal{I}_m$,
\begin{equation}
	\mathcal{I}[\rho]=\prod_m \mathcal{I}_m[\rho],
 \label{eq:qite_trotter}    
\end{equation}
effectively implements a weak projection onto the ground state manifold of $H$, as long as the the initial state has a nonzero overlap with this subspace.

In this work, as a target Hamiltonian $H$ we take the familiar one-dimensional transverse-field Ising chain:
\begin{equation}
    H = -J\sum_i\sigma^z_i \sigma^z_{i+1}+g\sum_{i}\sigma^x_i\equiv H_Z + H_X.
    \label{eq:TFIM}
\end{equation}
Here $J$ sets the overall energy scale and from now on we set $|J|=1$.
For both the ferromagnetic ($J=+1$) and the antiferromagnetic ($J=-1$) chain the Hamiltonian \eqref{eq:TFIM} possesses a $\mathbb{Z}_2$ symmetry generated by $U=\prod_i\sigma^x_i$, while its ground state exhibits an ordered (disordered) phase for transverse field values $g<1$ ($g>1$) and undergoes an SSB phase transition at the critical point $g=g_c=1$.

In the absence of noise and in the limit of small $\Delta\tau$, the steady state of the evolution~\eqref{eq:qite_trotter_m}--\eqref{eq:qite_trotter}
exhibits a (dynamical) phase transition inherited from the ground-state phase transition of $H$.
We now ask whether the ordered phase and the associated phase transition survive in the presence of noise.
We describe a noise channel affecting the circuit in terms of its Kraus operators $K_k$:
\begin{equation}
	\mathcal N[\rho]=\sum_k K_k\rho K_k^{\dagger},
 \label{eq:noise_op}
\end{equation}
where the Kraus operators satisfy the condition $\sum_k K_k^{\dagger} K_k=I$ and $I$ is the identity operator.
The specific form of the Kraus operators depends on the type of noise.
In this work we consider multiple types of noise: single-qubit bit-flip, depolarizing, and amplitude damping noise, as well as correlated two-qubit bit-flip noise.

\begin{figure}
	\includegraphics[height = 0.41\columnwidth]{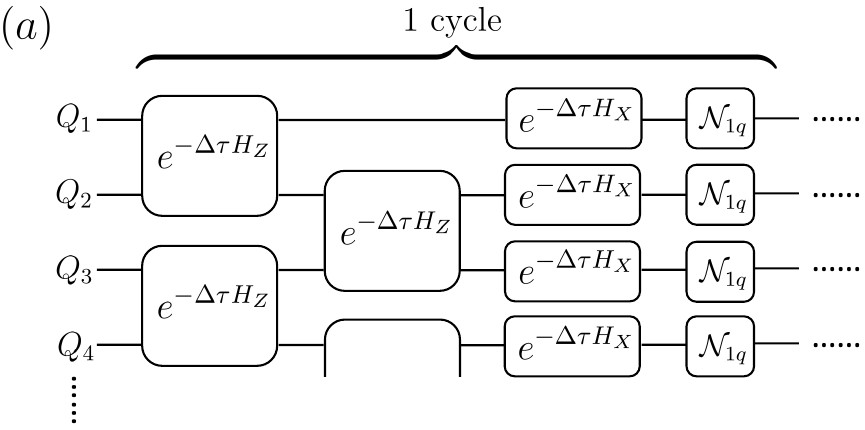}\par\bigskip
    \includegraphics[height = 0.41\columnwidth]{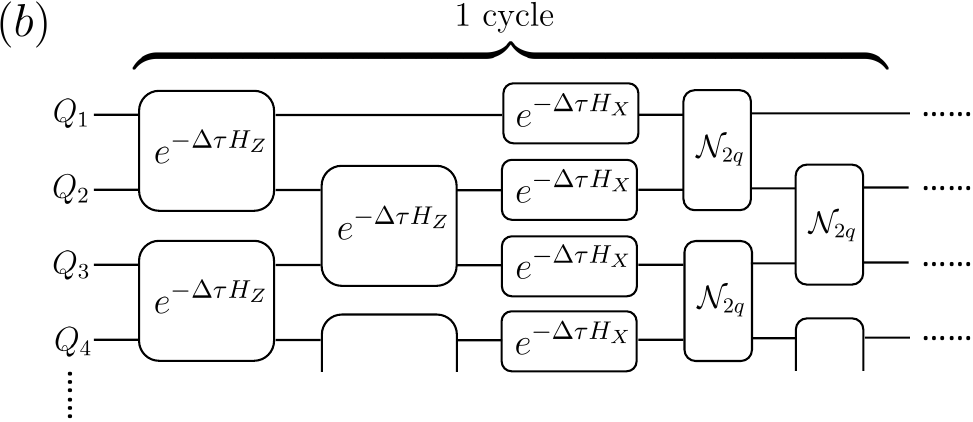}
	\caption{Schematic depiction of the circuit implementation of the noisy ITE evolution \eqref{eq:qite_noise} for the transverse-field Ising chain in the case of (a) single-qubit and (b) two-qubit noise. Here $e^{-\Delta\tau H_Z}$ and $e^{-\Delta\tau H_X}$ represent the Trotterized ITE parts of the protocol, while $\mathcal{N}_{1q}$ and $\mathcal{N}_{2q}$ denote the single- and two-qubit noise, respectively.}
	\label{fig:setup}
\end{figure}

The evolution that we investigate can be described as a joint action of the ITE and the noise operations at each Trotter step:
\begin{equation}
    \rho_{\rm st.s.}=\dots\mathcal{N}[\mathcal{I}[\mathcal{N}[\mathcal{I}[\rho_0]]]]\dots.
 \label{eq:qite_noise}  
\end{equation}
Here $\rho_0$ is some initial state that has a nonzero overlap with the ground state manifold of $H$, and the evolution \eqref{eq:qite_noise} is performed until the steady state $\rho_{\rm st.s.}$ is reached.
For the Ising chain \eqref{eq:TFIM} we schematically depict the ``circuit" implementation of one cycle of the evolution \eqref{eq:qite_noise} in Fig.~\ref{fig:setup}, where Figs.~\ref{fig:setup}(a) and (b) correspond to the cases of one-qubit and two-qubit noise, respectively.

To probe the steady state of the noisy ITE we employ the vectorization technique (also known as the Choi-Jamio\l kowski isomorphism \cite{Jamiolkowski1972,Choi1975}), rewriting the density operator $\rho$ as a state vector in a doubled Hilbert space.
We denote this state vector by $|\rho\rangle\!\rangle$, and define it via
\begin{equation}
    |\rho\rangle\!\rangle = \sum_{mn}\rho_{mn}|m\rangle\otimes|n\rangle,
\end{equation}
where $\rho_{mn}$ are matrix elements of the density operator $\rho = \sum_{mn}\rho_{mn}|m\rangle\langle n|$ in the computational basis of the original Hilbert space.
Within this formalism, the ITE superoperator $\mathcal{I}_m$ of Eq.~\eqref{eq:qite_trotter_m} becomes an operator acting on the doubled space:
\begin{equation}
    \tilde{\mathcal{I}}_m  \propto e^{-\Delta\tau h_m}\otimes (e^{-\Delta\tau h_m})^{\ast},
    \label{eq:ite_superoperator}
\end{equation}
where we utilize a tilde to denote operators in the doubled space and where we have omitted a $|\rho\rangle\!\rangle$-dependent normalization.
The noise superoperator $\mathcal{N}$ of Eq.~\eqref{eq:noise_op}, in turn, corresponds to an operator
\begin{equation}
    \tilde{\mathcal{N}}= \sum_k K_k\otimes K_k^{\ast}.
    \label{eq:noise_superoperator_Kraus_general}
\end{equation}
Given an (in general unnormalized) density operator $\rho$, the mixed state expectation value of an observable $O$ can be computed within the doubled space formalism as
\begin{equation}
    \langle O\rangle = \frac{\Tr[\rho O]}{\Tr[\rho]} = \frac{\langle\!\langle I| O\otimes I|\rho\rangle\!\rangle}{\langle\!\langle I|\rho\rangle\!\rangle},
    \label{eq:obs_general}
\end{equation}
where the state vector $|I\rangle\!\rangle=\sum_m|m\rangle\otimes|m\rangle$ corresponds (up to normalization) to the maximally mixed state.

\section{\label{sec:Results}Results}

Our analysis hinges on mapping the noisy ITE circuit \eqref{eq:qite_noise} to a noiseless ITE in the doubled Hilbert space, which can be treated efficiently using DMRG. We describe this mapping in detail for single-qubit bit-flip noise before proceeding to consider other noise channels. We focus on the ferromagnetic Ising model with $J=+1$, except in our discussion of amplitude damping noise in Sec.~\ref{sec:Amplitude_damping}.

\subsection{\label{sec:Pauli}Pauli noise channels}

\subsubsection{Single-qubit bit-flip noise}

We start by considering the ITE in the presence of single-qubit bit-flip errors (also known as $X$ noise) corresponding to the noise channel
\begin{align}
	&\mathcal N[\rho] =\prod_i\mathcal N_i[\rho],  \\
	&\mathcal N_i[\rho]=(1-p)\, \rho + p\, \sigma^x_i\rho \sigma^x_i,
\end{align}
where $i$ is the qubit index.
In the doubled space, this error channel takes the form
\begin{equation}
    \tilde {\mathcal N} =\prod_i [(1-p) + p\,\sigma^x_i\otimes \sigma^x_i] = e^{\mu\sum_i(\sigma^x_i\otimes \sigma^x_i-I\otimes I)},
    \label{eq:X_noise_superoperator}
\end{equation}
where in the second equality we have used the fact that $e^{\mu \sum_i \sigma^x_i\otimes \sigma^x_i}=\prod_i e^{\mu \sigma^x_i\otimes \sigma^x_i}=\prod_i[\cosh\mu+\sinh\mu \, \sigma^x_i\otimes \sigma^x_i]$ and have introduced a parameter $\mu=\tanh^{-1}[p/(1-p)]$, such that $e^{-\mu}\cosh\mu=1-p$ and $e^{-\mu}\sinh\mu=p$.
This parameter is real and positive for $0<p<1/2$ and in the limit of weak noise, $p\ll1$, becomes $\mu\sim p+O(p^2)$.
Given Eqs.~\eqref{eq:ite_superoperator} and \eqref{eq:X_noise_superoperator}, in the doubled space the steady state of the noisy ITE can be written, up to an overall normalization prefactor, as 
\begin{equation}
	|\rho_{\rm st.s.}\rangle\!\rangle\propto \left[e^{\mu \sum_i (\sigma^x_i\otimes \sigma^x_i)}\prod_m\left[e^{-\Delta\tau h_m}\otimes (e^{-\Delta\tau h_m})^{\ast}\right]\right]^l|\rho_0\rangle\!\rangle,
 \label{eq:rho_s_st_X_general}
\end{equation}
where $l$ is the number of Trotter steps required to reach the steady state.
Taking the limit $\Delta\tau\to 0$, $\mu\to 0$ allows one to neglect the Trotter error in the expression \eqref{eq:rho_s_st_X_general}, which then can be simplified into
\begin{equation}
	|\rho_{\rm st.s.}\rangle\!\rangle\propto \left[e^{-\Delta\tau(H\otimes I + I\otimes H^{\ast} -\lambda \sum_i \sigma^x_i\otimes \sigma^x_i)}\right]^l|\rho_0\rangle\!\rangle.
  \label{eq:rho_s_st_X_continuum}
\end{equation}
Here $H=\sum_m h_m$ is the Hamiltonian governing the ITE, whereas the newly introduced parameter $\lambda=\mu/\Delta\tau$ in the case of weak noise describes the error rate per Trotter step, $\lambda\approx p/\Delta\tau$.
Since noise in state of the art quantum devices can generally be considered weak, $p\ll 1$ \cite{quantum_supremacy19,morvan2023_pht_rcs_arxiv,xiang2024_topo_time_crystal_exp_arxiv}, hereafter we will refer to $\lambda$ as an effective error rate.

Equation~\eqref{eq:rho_s_st_X_continuum} shows that in the limit $\Delta\tau,p\to 0$, $|\rho_{\rm st.s.}\rangle\!\rangle$ is the steady state of the ITE with an effective Hamiltonian $\tilde{H}_{\rm eff}$ corresponding to two copies of the original system coupled by a noise-induced term $\tilde{H}_{\mathcal N}$:
\begin{align}
    &|\rho_{\rm st.s.}\rangle\!\rangle\propto \left[e^{-\Delta\tau \tilde{H}_{\rm eff}}\right]^l|\rho_0\rangle\!\rangle, \\
    &\tilde{H}_{\rm eff}=H\otimes I + I\otimes H^{\ast}+\tilde{H}_{\mathcal N}, \label{eq:H_eff_X_general} \\
    &\tilde{H}_{\mathcal N}=-\lambda\sum_i\sigma^x_{i}\otimes\sigma^x_{i}. \label{eq:H_12_X}
\end{align}
Consequently, $|\rho_{\rm st.s.}\rangle\!\rangle$ corresponds to the ground state of $\tilde{H}_{\rm eff}$, and therefore the steady state properties of the noisy ITE, such as the SSB order, can be obtained by studying the ground state of $\tilde{H}_{\rm eff}$.

When the ITE is governed by the transverse-field Ising chain Hamiltonian~\eqref{eq:TFIM}, $H=H^{\ast}$ and $\tilde H_{\rm eff}$ can be viewed as a transverse-field Ising ladder whose two legs are coupled by a noise-induced $XX$ term on the rungs.
For both the ferromagnetic and the antiferromagnetic model the $\mathbb Z_2$ symmetry generated by $U=\prod_i\sigma^x_i$ leads to a $\mathbb{Z}_2\times\mathbb{Z}_2$ symmetry of the effective Hamiltonian generated by $U\otimes I$ and $I\otimes U$.
This $\mathbb{Z}_2\times\mathbb{Z}_2$ symmetry corresponds to nothing but the strong (also known as  exact) symmetry of the associated noisy evolution~\cite{Buca_2012_strong_weak_sym, deGroot2022symmetryprotected}.
The ground state of the Hamiltonian \eqref{eq:H_eff_X_general}-\eqref{eq:H_12_X} thus possesses the $\mathbb{Z}_2\times\mathbb{Z}_2$ symmetry as well, unless this symmetry is spontaneously broken.
However, it is crucial to emphasize that when this ground state is utilized to calculate the mixed steady state observables via Eq.~\eqref{eq:obs_general}, the $\mathbb{Z}_2\times\mathbb{Z}_2$ symmetry breaks down to a global $\mathbb{Z}_2$ subgroup generated by $U\otimes U$ due to the overlap with the state $|I\rangle\!\rangle$.
This global $\mathbb{Z}_2$ symmetry corresponds to applying the original $\mathbb Z_2$ symmetry generator to both legs of the ladder simultaneously and represents the weak (or average) symmetry of the noisy evolution.
Hence, when observables \textit{linear} in the density matrix are calculated, only the weak symmetry is important and can be spontaneously broken, as the strong symmetry in this case is broken by default.
The situation is different when one is interested in quantities \textit{nonlinear} in the density matrix, such as the R\'enyi-2 correlator, for which the presence of the strong symmetry is relevant and can lead to various interesting physical phenomena, such as strong-to-weak SSB \cite{Ma2024topological_arxiv,Lessa2024strongtoweak_arxiv,Lee2023_criticality_under_decoherence,ma2024symmetry_arxiv,sala2024spontaneous_arxiv}.

The SSB of the weak $\mathbb Z_2$ symmetry generated by $U\otimes U$ leads to the ordered phase in the steady state of the noisy ITE.
Since the noise-induced term \eqref{eq:H_12_X} does not break the weak symmetry (in fact, by itself it even preserves the strong symmetry), we expect the ground state order and phase transition of the Ising chain \eqref{eq:TFIM} to persist in the steady state of the noisy ITE, as long as the noise is not too strong and the Trotter error is not too large.
Furthermore, since the relevant symmetry of the noisy problem is the same as that of the noiseless one, we expect its universality class to be the same as in the noiseless case (where the transition is in the 2D Ising universality class).
The location of the critical point, on the other hand, is nonuniversal and expected to shift in the presence of noise.

\begin{figure*}
	\includegraphics[width = 2.08\columnwidth]{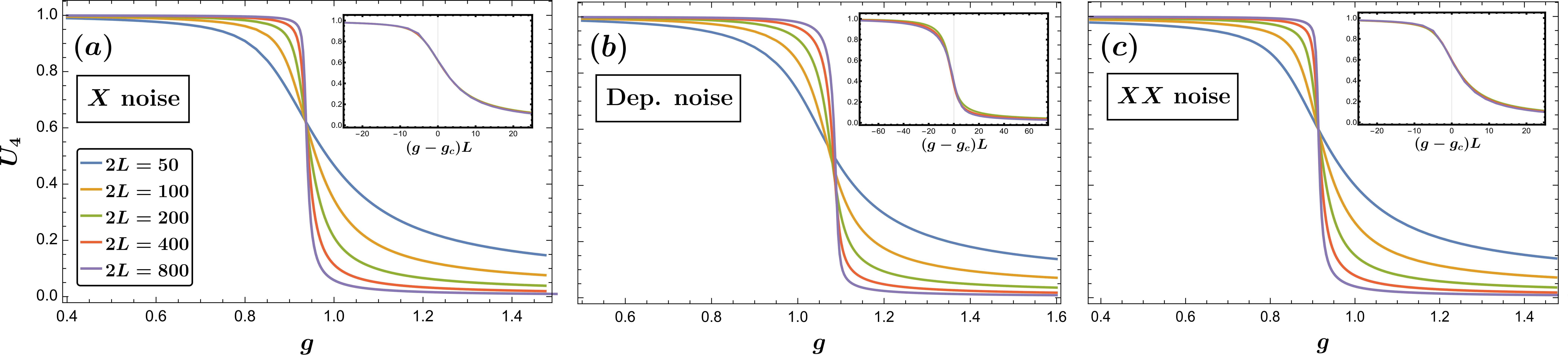}
	\caption{Binder cumulant from DMRG plotted as a function of the transverse field for the ferromagnetic ($J=+1$) Ising spin ladder \eqref{eq:H_eff_X_general} with interleg couplings induced by (a) single-qubit bit-flip ($X$) noise given by Eq.~\eqref{eq:H_12_X} with $\lambda=0.1$, (b) single-qubit depolarizing noise given by Eq.~\eqref{eq:H_12_XYZ} with $\lambda=0.15$, and (c) two-qubit bit-flip ($XX$) noise given by Eq.~\eqref{eq:H_12_XX} with $\lambda=0.1$. The mixed-state expectation values $\langle m^2 \rangle$ and $\langle m^4 \rangle$ entering the expression \eqref{eq:binder_cumulant} for the Binder cumulant are computed using Eq.~\eqref{eq:obs_general}. The curves are plotted for various ladder lengths $L$ in order to identify the scale invariant critical point $g_c$; the total number of sites in each DMRG calculation is equal to $2L$. The inset in each subfigure depicts the Binder cumulant as a function of $(g-g_c)L$ and demonstrates the collapse of the curves corresponding to different system sizes. Open boundary conditions are assumed throughout the DMRG calculations.}
	\label{fig:dmrg_Pauli_noise}
\end{figure*}

To showcase the above statements, we numerically investigate~\cite{ite_noise_code_data} the symmetry breaking order-disorder transition in the steady state of the noisy ITE by computing the ground state of the spin ladder Hamiltonian \eqref{eq:H_eff_X_general}-\eqref{eq:H_12_X} using the DMRG \cite{White1992dmrg,White1993dmrg,Schollwock2005_dmrgRMP} and then calculating the expectation values of observables in the steady state via Eq.~\eqref{eq:obs_general}.
Numerical details of the DMRG calculations are presented in Appendix~\ref{sec:dmrg_details}.
To identify the location of the transition as precisely as possible, we calculate the Binder cumulant~\cite{Binder1981FiniteSS, Binder1981CriticalPF},
\begin{equation}
    U_4=\frac{1}{2}\left(3-\frac{\langle m^4 \rangle}{\langle m^2 \rangle^2}\right),
    \label{eq:binder_cumulant}
\end{equation}
where $m$ is the order parameter, which for the case of the $L$-site ferromagnetic Ising chain is the average magnetization $m=\sum_i\sigma_i^z/L$.
The Binder cumulant \eqref{eq:binder_cumulant} exhibits scale invariance at the critical point $g=g_c$ \cite{sandvik2010}, whereas near the critical point it is expected to scale with the system size as $U_4(g,L)=f[(g-g_c)L^{1/\nu}]$, which allows one to extract the correlation length critical exponent $\nu$.

Figure~\ref{fig:dmrg_Pauli_noise}(a) depicts the Binder cumulant evaluated for the DMRG-calculated ground state of the spin ladder Hamiltonian \eqref{eq:H_eff_X_general}--\eqref{eq:H_12_X} as a function of the transverse field $g$ for various values of the ladder length $L$ and open boundary conditions. 
The Ising couplings along the legs of the ladder are taken to be ferromagnetic, while the strength of the noise-induced couplings along the rungs is set to $\lambda=0.1$.
For this value of $\lambda$ the critical point, identified through the scale invariance of the Binder cumulant, is located at $g_c\approx 0.94$, which is slightly lower than the noiseless value of $g_c=1$.
The ordered phase is thus still present in the noisy model, although its extent in parameter space is smaller when compared to the noiseless case.
The curves representing different system sizes collapse when the Binder cumulant is plotted as a function of $(g-g_c)L$, which is illustrated by the inset in Fig.~\ref{fig:dmrg_Pauli_noise}(a).
The finite size scaling thus yields the correlation length critical exponent of $\nu=1$, which, as expected, belongs to the 2D Ising universality class.
In Appendix~\ref{sec:gap_CF} we also plot the dependence of the energy gap on $g-g_c$ near critical point, from which we extract dynamical critical exponent $z=1$, and showcase power-law decay of the $\langle\sigma^z_i\sigma^z_j\rangle$ correlation function at criticality, extracting critical exponent $\eta=1/4$.
The values of these critical exponents are consistent with the 2D Ising universality class.

\begin{figure}
	\includegraphics[width = 0.95\columnwidth]{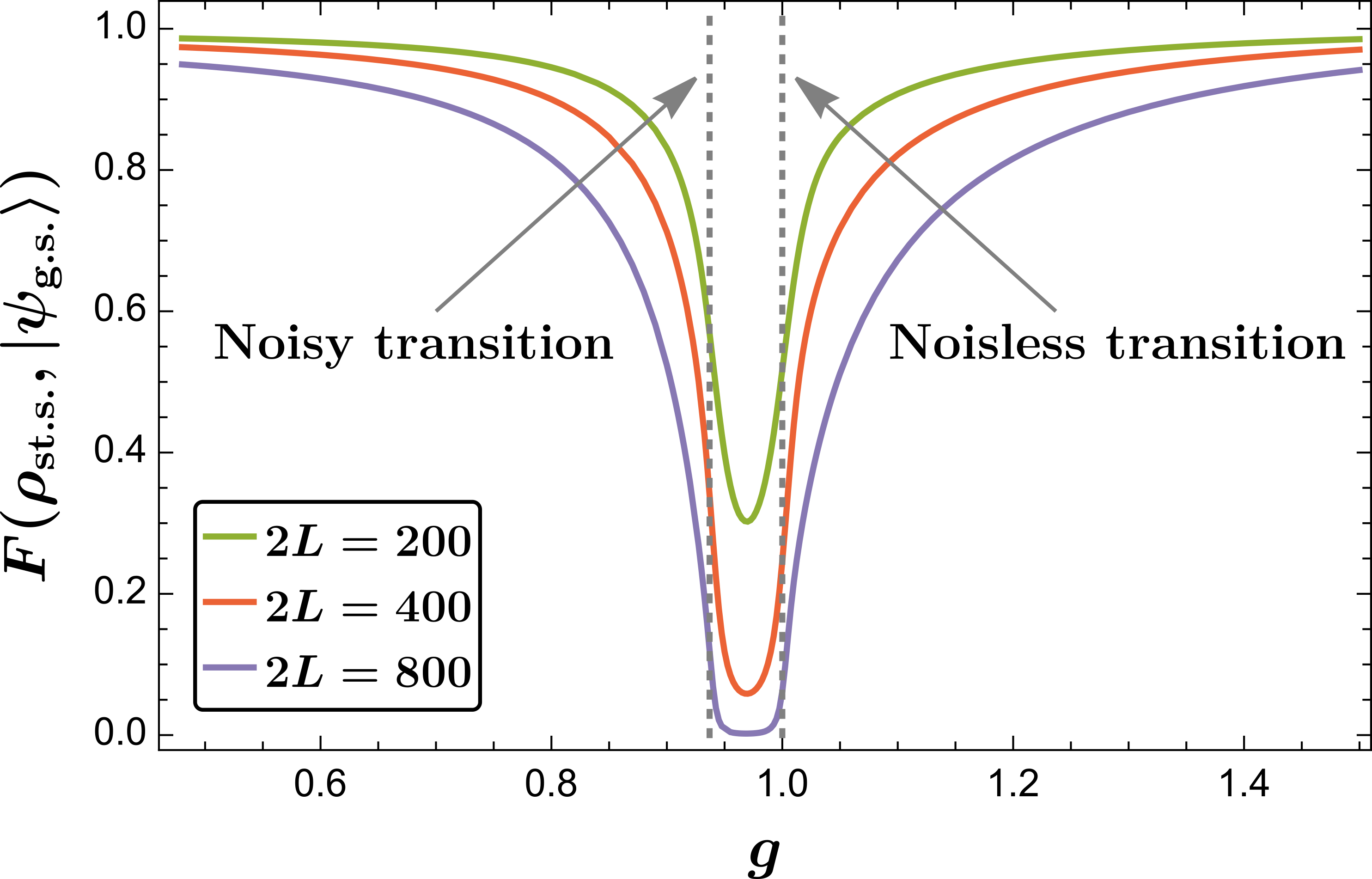}
	\caption{Fidelity between the mixed noisy steady state $\rho_{\rm st.s.}$ and the pure target ground state $|\psi_{\rm g.s.}\rangle$ as a function of the transverse field. The fidelity is computed using the DMRG with open boundary conditions via Eqs.~\eqref{eq:fidelity}--\eqref{eq:rho_overlap}. The noise-induced coupling in the effective Hamiltonian is set to $\lambda=0.1$. The dotted lines indicate the critical points for the SSB transition in the noisy ($g_c\approx0.94$) and noiseless ($g_c=1$) steady state. A small symmetry-breaking field is added at the ends of the spin ladder to ensure that the DMRG algorithm finds a symmetry-broken ground state within the ground-state manifold.}
	\label{fig:fidelity}
\end{figure}

In addition to identifying the presence of the SSB order in the noisy steady state $\rho_{\rm st.s.}$, we also quantify how ``close" this state is to the target state, namely the ground state $|\psi_{\rm g.s.}\rangle$ of the original Hamiltonian $H$.
We quantify this with the fidelity,
\begin{equation}
    F(\rho_{\rm st.s.},|\psi_{\rm g.s.}\rangle) = \langle \psi_{\rm g.s.}| \rho_{\rm st.s.}|\psi_{\rm g.s.}\rangle = \Tr[\rho_{\rm st.s.}\rho_{\rm g.s.}],
    \label{eq:fidelity}
\end{equation}
where $\rho_{\rm g.s.}=|\psi_{\rm g.s.}\rangle  \langle \psi_{\rm g.s.}|$ is the pure state density matrix corresponding to the ground state of $H$.
Within the doubled space formalism, the trace in Eq.~\eqref{eq:fidelity} can be evaluated as an overlap between the ground states of the effective Hamiltonian with and without the noise-induced coupling:
\begin{equation}
    \Tr[\rho_{\rm st.s.}\rho_{\rm g.s.}] = \langle\!\langle \rho_{\rm g.s.} | \rho_{\rm st.s.}\rangle\!\rangle.
    \label{eq:rho_overlap}
\end{equation}
We compute the two ground states, one with $\lambda=0.1$ and the other with $\lambda=0$, and the overlap between them using DMRG, and plot the resulting fidelity as a function of the transverse field in Fig.~\ref{fig:fidelity}. 
For $g\ll1$ ($g\gg1$), both the noisy and the noiseless steady state are deep in their respective ordered (disordered) phases, and the fidelity between them is close to one.
On the other hand, for values of $g$ in between the noiseless and the noisy critical points (marked by dotted lines in the figure), the noisy steady state is disordered, while the noiseless one is ordered, and the fidelity dips drastically toward zero.
The dip is stronger for larger system sizes, with the fidelity reaching zero for $2L=800$.
We note that the fidelity also decreases with system size even deep in the ordered/disordered phase, albeit the decrease is much smaller there than in the region between the critical points.
We attribute this decrease to the orthogonality catastrophe and expect the fidelity to decay to zero exponentially with system size for any value of $g$ in the thermodynamic limit, $L\to\infty$.
While we observe exponential decay within the region between the noiseless and the noisy critical points, unfortunately, the available system sizes preclude us from definitively inferring exponential decay outside of this region.
We demonstrate scaling of the fidelity with system size in Appendix~\ref{sec:fidelity_syst_size}.

\begin{figure*}
	\includegraphics[width = 2.05\columnwidth]{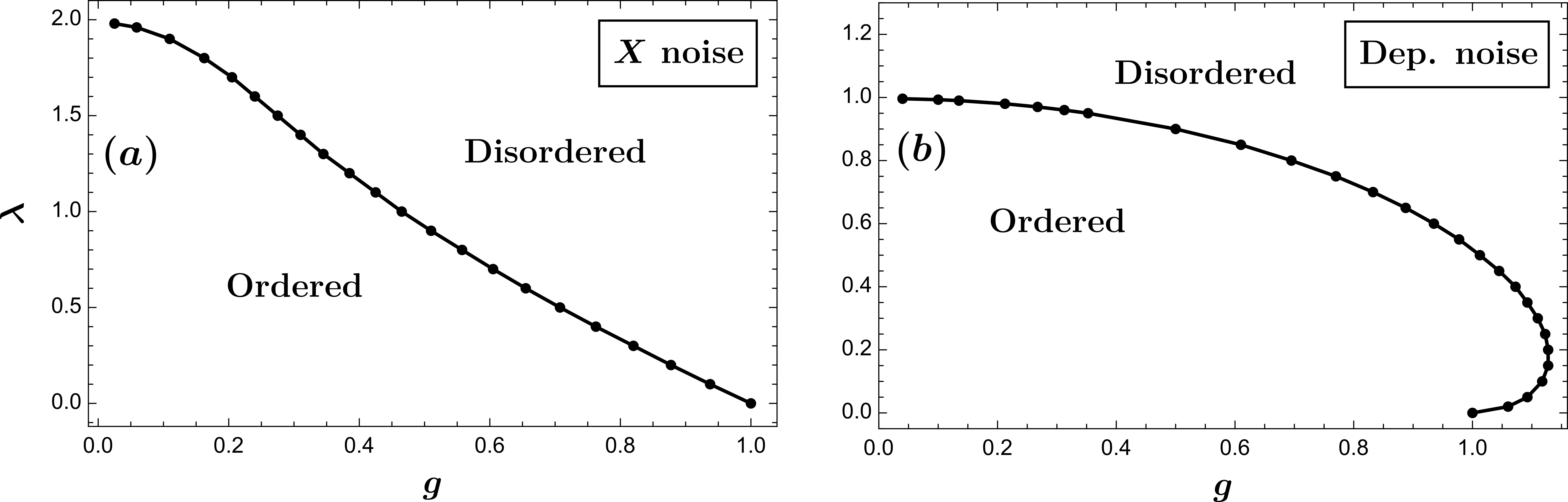}
	\caption{Phase diagram for the ferromagnetic ($J=+1$) Ising spin ladder model \eqref{eq:H_eff_X_general} when the interleg couplings are induced by (a) single-qubit bit-flip ($X$) noise given by Eq.~\eqref{eq:H_12_X} and (b) single-qubit depolarizing noise given by Eq.~\eqref{eq:H_12_XYZ}. The phase boundaries are obtained by computing the Binder cumulant \eqref{eq:binder_cumulant} using the DMRG and extracting the scale-invariant crossings between (a) the $2L=200$ and $2L=400$ curves and  (b) the $2L=400$ and $2L=800$ curves.}
	\label{fig:dmrg_phase_diagram}
\end{figure*}

Using the DMRG, we further calculate the location of the critical point $g_c$ for other values of the parameter $\lambda$ and plot the resulting two-dimensional phase diagram in the $(g,\lambda)$ parameter space in Fig.~\ref{fig:dmrg_phase_diagram}(a).
This phase diagram demonstrates that in the limit $\Delta\tau,p\to 0$ of the noisy ITE the ordered phase is stable in the presence of the single-qubit bit-flip noise, and survives as long as the effective error rate $\lambda=p/\Delta\tau$ is below the upper critical value of $\lambda_{c,\rm max}\approx 2$. This calculation therefore validates the schematic picture shown in Fig.~\ref{fig:schematic}.

While we have numerically shown the stability of the ordered phase against the $X$ noise in the limit where both the Trotter step and the noise strength are small $(\Delta\tau,p\to 0)$ such that the effective Hamiltonian picture \eqref{eq:H_eff_X_general}--\eqref{eq:H_12_X} applies, an actual implementation of the ITE on quantum hardware would have a finite Trotter step and a finite noise strength.
To investigate whether the SSB ordered phase and the phase transition are still present in that latter scenario, we numerically simulate~\cite{ite_noise_code_data} the density matrix throughout the noisy ITE process for small system sizes on a classical high-performance computer with graphics processing units (GPUs).
The simulated circuit is schematically depicted in Fig.~\ref{fig:setup}(a).

\begin{figure*}
	\includegraphics[width = 2.05\columnwidth]{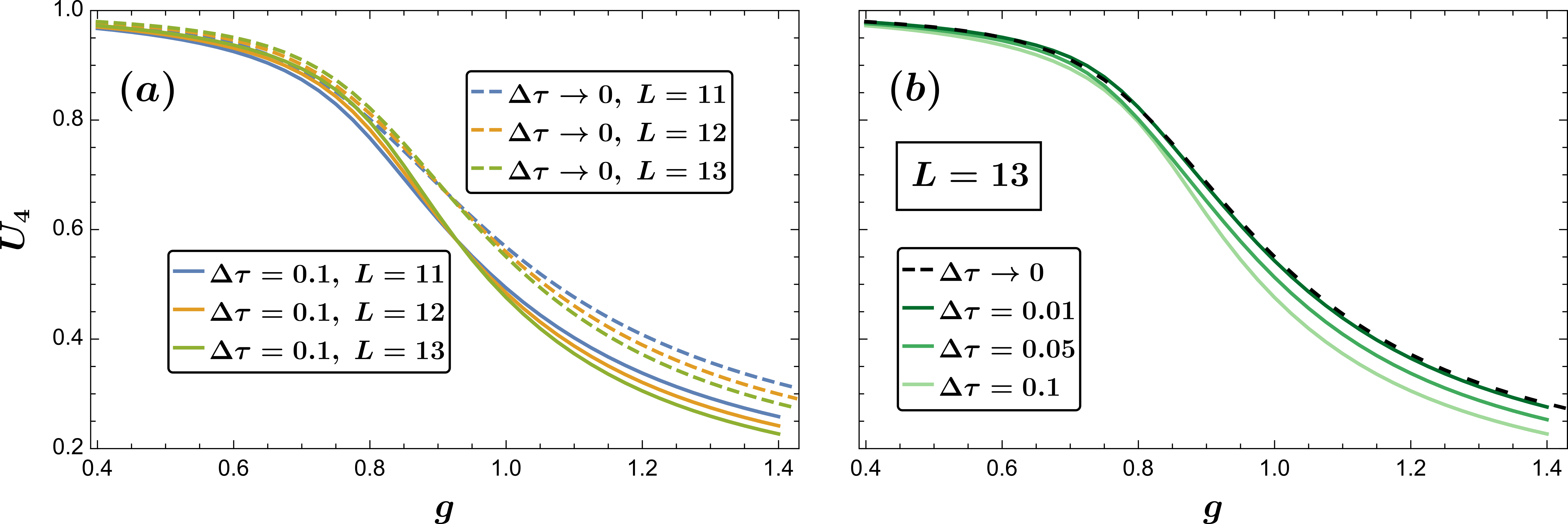}
	\caption{(a) Solid lines: Binder cumulant plotted as a function of the transverse field for the steady state of a directly simulated density matrix throughout the noisy ITE process with ferromagnetic couplings ($J=+1$), $X$ noise and a finite ($\Delta\tau=0.1$) Trotter step. Ferromagnetically fully ordered state is taken as the initial state for the noisy ITE simulations. Dashed lines: same curves but plotted for the DMRG-calculated ground state of the Ising spin ladder model \eqref{eq:H_eff_X_general} with the $X$-noise-induced interleg couplings \eqref{eq:H_12_X}. The dashed lines represent the results for the noisy ITE with a vanishing Trotter step ($\Delta\tau\to0$). In both cases the curves are plotted for various spin chain/ladder lengths $L$ in order to identify the scale invariant point. The effective error rate in both cases is set to $\lambda=0.1$ and open boundary conditions are assumed. (b) Same curves but plotted for a fixed length of the spin chain/ladder ($L=13$) and various values of the ITE Trotter step.}
	\label{fig:dmrg_DMev}
\end{figure*}

The results of these density matrix simulations are represented by solid curves in Fig.~\ref{fig:dmrg_DMev}(a), which, similarly to Fig.~\ref{fig:dmrg_Pauli_noise}(a), depicts the Binder cumulant as a function of the transverse field for various lengths of the chain.
The Binder cumulant in this case is calculated for the steady state of the noisy ITE process with a finite Trotter step $\Delta\tau=0.1$ and the effective error rate of $\lambda=p/\Delta\tau=0.1$.
When simulating the noisy ITE, we take ferromagnetically fully ordered state $|\uparrow\uparrow\dots\uparrow\rangle$ as the initial state, but we checked that starting with an antiferromagnetically ordered or a disordered state yields the same results.
The dashed curves in Fig.~\ref{fig:dmrg_DMev}(a) on the other hand represent the Binder cumulant calculated from DMRG on the effective Hamiltonian \eqref{eq:H_eff_X_general}--\eqref{eq:H_12_X}, and thus represent the limit $\Delta\tau,p\to 0$ of the noisy ITE with the same $\lambda=p/\Delta\tau=0.1$ \footnote{The finite size crossing point here is slightly shifted from the value obtained from the large-system-size curves [depicted in Fig.~\ref{fig:dmrg_Pauli_noise}(a)] due to a finite size drift caused by subleading terms in the Binder cumulant expansion near the critical point.}.
The scale invariant crossing of the solid curves in Fig.~\ref{fig:dmrg_DMev}(a) illustrates that the ordered phase and the phase transition are present in the noisy ITE even if $\Delta\tau$ is finite.
Still, finite Trotter error slightly shifts the curves and the scale invariant point from the results in the limit $\Delta\tau,p\to 0$.
This shift can be systematically reduced by reducing the Trotter step size.
We demonstrate this by plotting the Binder cumulant versus transverse field curves for various values of the Trotter step, ranging from $\Delta\tau=0.01$ to $\Delta\tau=0.1$, together with the $\Delta\tau\to0$ results, in Fig.~\ref{fig:dmrg_DMev}(b).
Here we fix the effective error rate to $\lambda=p/\Delta\tau=0.1$ and the chain length to $L=13$ sites.
These results validate the reliability of the effective Hamiltonian approach in the limit of weak noise.
Given the advantages of the DMRG approach in terms of scalability and mitigating finite-size effects, we therefore restrict our focus to this approach for the remainder of the paper.

\subsubsection{Single-qubit depolarizing noise}

Next, we study ITE in the presence of the single-qubit depolarizing noise.
This noise channel reads:
\begin{align}
	&\mathcal N[\rho] =\prod_i\mathcal N_i[\rho],  \\
	&\mathcal N_i[\rho]=(1-p)\, \rho + \frac{p}{3}\, \sigma^x_i\rho \sigma^x_i + \frac{p}{3}\, \sigma^y_i\rho \sigma^y_i + \frac{p}{3}\, \sigma^z_i\rho \sigma^z_i,
\end{align}
where $i$ is again the qubit index.
For the depolarizing noise one can also derive an effective Hamiltonian in the doubled Hilbert space, the ground state of which describes the mixed steady state of the noisy ITE in the $\Delta\tau,p\to 0$ limit.
This effective Hamiltonian also takes the form of a spin ladder, see Eq.~\eqref{eq:H_eff_X_general}, but with a different noise-induced coupling on the rungs:
\begin{align}
    \tilde{H}_{\mathcal N}=-\frac{\lambda}{3}\sum_i\left(\sigma^x_{i}\otimes\sigma^x_{i}-\sigma^y_{i}\otimes\sigma^y_{i}+\sigma^z_{i}\otimes\sigma^z_{i} \right),
    \label{eq:H_12_XYZ}
\end{align}
where the minus sign in the $\sigma^y_{i}\otimes\sigma^y_{i}$ term is present due to the complex conjugation appearing in Eq.~\eqref{eq:noise_superoperator_Kraus_general}.
Here we have introduced $\lambda=\mu/\Delta\tau$ and $\mu = -3/4\log(1-4p/3)$; the parameter $\mu$ is real and positive for $0<p<3/4$.
Akin to the case of the bit-flip error, here the weak noise limit yields $\mu\sim p+O(p^2)$ and thus in this limit the parameter $\lambda$ describes the effective error rate per the ITE Trotter step, $\lambda\approx p/\Delta\tau$.

In the ferromagnetic case, the Hamiltonian \eqref{eq:H_12_XYZ} breaks the strong $\mathbb{Z}_2\times\mathbb{Z}_2$ symmetry, but preserves the weak $\mathbb{Z}_2$ symmetry generated by the simultaneous bit-flip on both chains, which is the relevant symmetry when calculating the mixed steady-state observables linear in the density matrix.
Hence, we expect the SSB order and the phase transition in the steady state of the ITE to be stable in the presence of the depolarizing noise, and we further expect the noisy transition to belong to the same 2D Ising universality class.

To numerically confirm these expectations, similarly to the $X$-noise spin ladder \eqref{eq:H_eff_X_general}--\eqref{eq:H_12_X}, here we calculate the ground state of the depolarizing-noise spin ladder \eqref{eq:H_eff_X_general},\eqref{eq:H_12_XYZ} with DMRG and plot the Binder cumulant extracted for that ground state as a function of the transverse field for various system sizes in Fig.~\ref{fig:dmrg_Pauli_noise}(b).
The interleg coupling is set to $\lambda=0.15$.
As depicted in Fig.~\ref{fig:dmrg_Pauli_noise}(b), for this value of $\lambda$ the almost-scale-invariant crossing point of curves for different system sizes is shifted from a noiseless value of $g=1$ to a value $g>1$, in contrast to the model with $X$ noise where the shift was toward a value of $g<1$.
The ``almost" scale invariance of the crossing and its size-dependent drift can be explained by the presence of finite-size corrections to the Binder cumulant near the critical point.
In this work, the maximum system size that we numerically analyze is $2L=800$, and the crossing between the $2L=800$ and $2L=400$ curves for $\lambda=0.15$ occurs at $g_c\approx 1.09$.
However, the finite-size drift suggests that the actual critical point is located at a slightly larger value of $g$.
Notably, this drift appears stronger for the depolarizing noise model than for the bit-flip noise model; this can be seen by examining the inset of Fig.~\ref{fig:dmrg_Pauli_noise}(b), which depicts the Binder cumulant as a function of $(g-g_c)L$.
While the inset curves do collapse close to each other, suggesting the same correlation length critical exponent of $\nu=1$ as in the case of the bit-flip noise, the collapse is not as dense as the one depicted in the inset of Fig.~\ref{fig:dmrg_Pauli_noise}(a).

By calculating the crossing points between the $2L=400$ and $2L=800$ curves for various values of $\lambda$, we obtain an approximate phase diagram in the $(g,\lambda)$ plane for the steady state of the noisy ITE with the depolarizing noise in the $\Delta\tau,p\to 0$ limit.
This phase diagram is plotted in Fig.~\ref{fig:dmrg_phase_diagram}(b).
We emphasize that the computed phase boundary is approximate due to the aforementioned finite-size drift of the crossing, but we expect the true phase boundary to lie close to the one we obtained. 

In contrast to the phase diagram for the bit-flip noise in Fig.~\ref{fig:dmrg_phase_diagram}(a), here the critical $g_c$, and with it the extent of the ordered phase, first increases as $\lambda$ is raised.
Past a certain value of $\lambda$, which appears to be close to $\lambda=0.2$, the critical value of $g_c$ starts to decrease.
Consequently, and perhaps contrary to naive expectations, the maximal extent of the ordered phase in the $g$ parameter space is reached at a nonzero level of noise.
We speculate that this phenomenon can be attributed to the fact that $\sigma^z_i$ operators within the depolarizing noise channel anticommute with $\sigma^x_i$ operators within the $e^{-\Delta\tau g \sigma^x_i}$ part of an ITE step, which may, for a small level of noise and after averaging over different quantum trajectories, effectively reduce the value of the transverse field felt by the system, thus requiring larger values of $g$ to move out of the ordered phase.
Overall, Fig.~\ref{fig:dmrg_phase_diagram}(b) illustrates that similarly to the case of $X$ noise, the ITE in the presence of the single-qubit depolarizing noise boasts a stable SSB ordered phase, and the order survives up to a maximum effective error rate of $\lambda_{c,\rm max}\approx 1$.
This example also highlights that it is the weak $\mathbb Z_2$ rather than the strong $\mathbb Z_2\times\mathbb Z_2$ symmetry that protects the SSB transition in the noisy ITE.

\subsubsection{Two-qubit bit-flip noise}

We now investigate ITE in the presence of a two-qubit noise.
As a particular example, we take the two-qubit bit-flip ($XX$) noise, the error channel for which can be written as
\begin{align}
	&\mathcal N[\rho] =\prod_i\mathcal N_{i,i+1}[\rho],  \\
	&\mathcal N_{i,i+1}[\rho]=(1-p)\, \rho + p\, \sigma^x_i\sigma^x_{i+1}\rho \sigma^x_i\sigma^x_{i+1} ,
\end{align}
where $i$ is the qubit index.
This noise channel induces correlated errors, whereby a bit flip on site $i$ is always accompanied by a bit flip on site $i+1$.
The effective Hamiltonian in the doubled Hilbert space in this case can be obtained in exactly the same fashion as the one for the ITE with the single-qubit bit-flip noise.
The effective Hamiltonian is again an Ising spin ladder, but with plaquette-type interleg coupling:
\begin{equation}
    \tilde{H}_{\mathcal N}=-\lambda\sum_i\sigma^x_{i}\sigma^x_{i+1}\otimes\sigma^x_{i}\sigma^x_{i+1}.
    \label{eq:H_12_XX}
\end{equation}
Here, as before, $\lambda=\mu/\Delta\tau$, $\mu=\tanh^{-1}[p/(1-p)]$, and in the weak noise limit $\lambda \approx p/\Delta\tau$ represents the effective error rate.
Note that in the ferromagnetic case the Hamiltonian \eqref{eq:H_12_XX} preserves the weak $\mathbb{Z}_2$ symmetry generated by the simultaneous bit-flip on both legs of the ladder.
Therefore, similarly to the single-qubit noise types considered before, here we expect the SSB order and associated transition to be stable and to remain within the 2D Ising universality class.

The Binder cumulant computed from DMRG on the effective Hamiltonian~\eqref{eq:H_eff_X_general},\eqref{eq:H_12_XX} with interleg coupling $\lambda=0.1$ is shown in Fig.~\ref{fig:dmrg_Pauli_noise}(c).
Overall, the behavior of the Binder cumulant in this case is very similar to the one presented for the case of the single-qubit bit-flip noise in the beginning of this section.
The scale-invariant crossing of the curves corresponding to different lengths of the spin ladder is located at $g_c\approx 0.91$, indicating the presence of the critical point and the SSB phase transition.
The Binder cumulant, plotted as a function of $(g-g_c)L$ in the inset of the figure, shows the collapse of the curves corresponding to different system sizes and suggests a correlation length critical exponent of $\nu=1$. 
Note that for this type of noise the finite-size drift of the crossing is more akin to the (very slight) drift occurring in the case of the $X$-noise and is not as strong as that observed for the case of the depolarizing noise.
We opt out of presenting the full phase diagram in the $(g,\lambda)$ plane for this type of noise since it very closely resembles the already computed phase diagram for the single-qubit bit-flip noise, see Fig.~\ref{fig:dmrg_phase_diagram}(a), indicating that the SSB ordered phase is stable under the noisy ITE even in the presence of correlated two-qubit noise, such as the $XX$ noise considered here.

In fact, we expect the phase transition to persist in the presence of \textit{any} Pauli noise. For any such noise, the interleg coupling in the effective Hamiltonian is of the form
\begin{align}
    \tilde H_{\mathcal N}=\sum_m c_m P_m\otimes P^*_m,
\end{align}
where $\{P_m\}$ are a set of Pauli strings and $c_m$ are real coefficients that depend on the noise strength. 
Since any Pauli string $P_m$ either commutes or anticommutes with the $\mathbb Z_2$ symmetry generator $U = \prod_i\sigma^x_i$, the weak symmetry generator $U\otimes U$ always commutes with $P_m\otimes P^*_m$. 

\subsection{\label{sec:Amplitude_damping}Amplitude damping noise channel}

\begin{figure*}
	\includegraphics[width = 2.05\columnwidth]{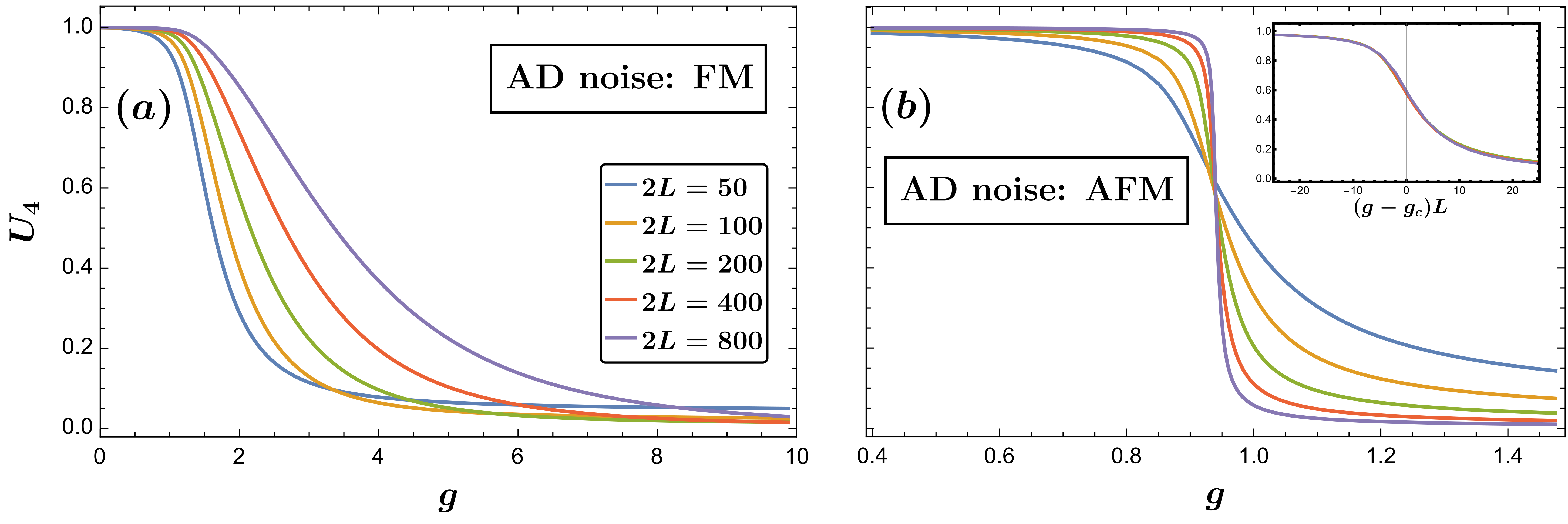}
	\caption{Binder cumulant plotted as a function of the transverse field and calculated using the DMRG for (a) the ferromagnetic (FM, $J=+1$) and (b) the antiferromagnetic (AFM, $J=-1$) Ising spin ladder model \eqref{eq:H_eff_X_general} when the interleg coupling is induced by the amplitude damping (AD) noise given by Eq.~\eqref{eq:H_12_AD} with $\lambda_z=\lambda_+=0.4$. The curves are plotted for various ladder lengths $L$ in order to identify the absence (in the FM case) or the presence (in the AFM case) of the scale invariant critical point $g_c$; the total number of sites in the DMRG calculation is equal to $2L$. The inset in (b) depicts the Binder cumulant as a function of $(g-g_c)L$ and demonstrates the collapse of the curves corresponding to different system sizes. Open boundary conditions are assumed throughout the calculations in both (a) and (b).}
	\label{fig:dmrg_AD}
\end{figure*}
We now move beyond Pauli noise by considering ITE under the amplitude damping (AD) noise channel.
The amplitude damping is an example of a non-unital quantum channel.
It models energy dissipation effects in open quantum systems~\cite{nielsen2002quantum}. 
In terms of Kraus operators, this channel can be written as
\begin{align}
	&\mathcal N[\rho] =\prod_i\mathcal N_i[\rho],  \\
	&\mathcal N_i[\rho]=K_{0,i}\rho K_{0,i}^{\dagger} + K_{1,i}\rho K_{1,i}^{\dagger},
\end{align}
where
\begin{align}
    &K_{0,i} = \frac{1+\sqrt{1-p}}{2}+ \frac{1-\sqrt{1-p}}{2}\, \sigma_i^z, \\
    &K_{1,i} = \sqrt{p}\, \sigma_i^+.
\end{align}
The effective Hamiltonian in the doubled space now corresponds to an Ising spin ladder with the following noise-induced term:
\begin{align}
    \tilde{H}_{\mathcal N}=-\frac{\lambda_z}{4}\sum_i \left(\sigma^z_{i}\otimes I+I\otimes\sigma^z_{i} \right)-\lambda_+\sum_i\sigma^+_{i}\otimes\sigma^+_{i},
    \label{eq:H_12_AD}
\end{align}
where $\sigma^+_{i}=(\sigma^x_{i}+i\sigma^y_{i})/2$ and we have introduced parameters $\lambda_{z,+}=\mu_{z,+}/\Delta\tau$ and $\mu_z=-\log(1-p),\ \mu_+=p$.
The limit of weak noise, $p\ll 1$, gives $\mu_z\approx p + O(p^2)$ and in this case both $\lambda_z$ and $\lambda_+$ represent the effective noise rate, $\lambda_{z,+}\approx p/\Delta\tau$.
The first sum in the Hamiltonian \eqref{eq:H_12_AD} represents a longitudinal field added to each leg of the Ising spin ladder, while the second sum is a non-Hermitian coupling along the rungs.
The entire effective Hamiltonian \eqref{eq:H_eff_X_general},\eqref{eq:H_12_AD} is thus non-Hermitian; however, being completely real it possesses a $\mathcal{PT}$ symmetry \cite{Bender1998_pt_sym,Bender1999_pt_sym,Bender2005_pt_sym_rev} and its eigenvalues are either real or come in complex-conjugated pairs.

If the Ising ladder has ferromagnetic couplings along its legs, the Hamiltonian \eqref{eq:H_12_AD}, in addition to breaking the strong $\mathbb{Z}_2\times\mathbb{Z}_2$ symmetry, also breaks the weak $\mathbb{Z}_2$ symmetry generated by the simultaneous bit-flip on both legs of the ladder.
Consequently, the ground state of the combined Hamiltonian \eqref{eq:H_eff_X_general},\eqref{eq:H_12_AD} is not expected to exhibit SSB.
We confirm this by calculating the Binder cumulant as a function of the transverse field using DMRG for various lengths of the spin ladder, see Fig.~\ref{fig:dmrg_AD}(a).
Note that the DMRG calculation here is done for a non-Hermitian Hamiltonian and yet yields a real ground state energy and observables, thus manifesting the aforementioned $\mathcal{PT}$ symmetry.
As opposed to the $U_4$ versus $g$ plots presented earlier (see Fig.~\ref{fig:dmrg_Pauli_noise}), here the curves corresponding to different system sizes intersect at different $g$ values, demonstrating the absence of a scale invariant crossing, and, hence, the phase transition.
This behavior of the Binder cumulant closely resembles the one occurring in the ferromagnetic mixed-field Ising chain, where it is well-known that the Ising symmetry and the SSB order are absent.
Given these results, we conclude that the SSB ordered phase and phase transition present in the steady state of noiseless ITE under the ferromagnetic quantum Ising Hamiltonian does not survive when the ITE is subject to AD noise.
In addition to computing $U_4$, we compute fidelity between the target ground state and the steady state of the noisy ITE in the presence of the AD noise using Eqs.~\eqref{eq:fidelity}-\eqref{eq:rho_overlap}.
These results are presented in Appendix~\ref{sec:fidelity_AD}.

The situation is drastically different for the antiferromagnetic Ising chain.
The order parameter in this case is the staggered magnetization $m_s=\sum_i(-1)^i\sigma_i^z/L$, which, when finite, indicates a breaking of both global spin-flip and mirror reflection symmetries.
In this case, explicitly breaking the global spin-flip symmetry by adding a uniform longitudinal field $h$ leaves the residual mirror symmetry intact, leading to a nontrivial phase diagram in the $(g,h)$ plane for the antiferromagnetic mixed-field Ising model~\cite{ovchinnikov2003}.
In the doubled space, the AD-noise-induced Hamiltonian \eqref{eq:H_12_AD} also preserves the relevant weak $\mathbb{Z}_2$ symmetry, which in this case is the mirror symmetry with respect to the plane perpendicular to the ladder, and thus the model \eqref{eq:H_eff_X_general},\eqref{eq:H_12_AD} should exhibit the SSB order and the phase transition.
This is demonstrated in Fig.~\ref{fig:dmrg_AD}(b), where as before we depict the Binder cumulant as a function of the transverse field for various ladder lengths and set $J=-1$.
This figure is similar to those presented for the other types of noise to which the transition is robust, see Fig.~\ref{fig:dmrg_Pauli_noise}. 
The Binder cumulant displays a scale-invariant crossing at $g_c\approx 0.94$.
The inset displays the collapse of the curves as a function of $(g-g_c)L$ and indicates the expected critical exponent of $\nu=1$.
These results further corroborate our main conclusion that the SSB ordered phase in the steady state of the noisy ITE is stable against noise as long as the noise preserves the underlying weak symmetry of the evolution.

\section{\label{sec:Conclusion}Conclusion}

In this paper we studied whether ITE-based preparation of SSB ground state order on a quantum computer is robust to noise.
For the sake of generality and simplicity, we did not consider any particular ITE algorithm, instead focusing on a black box approach representing ITE as a Trotterized imaginary-time ``circuit" and investigating the SSB in the mixed steady state of that circuit.
The main conclusion of our work is that, given a target Hamiltonian with a ground state that spontaneously breaks a symmetry, the preparation of the ground state order using the ITE is robust against noise, as long as the noise channel respects the weak version of that symmetry.

To illustrate this conclusion, we employed the Choi-Jamio\l kowski isomorphism and in the limit of weak noise and small imaginary-time Trotter step mapped the noisy ITE to a noiseless ITE under an effective Hamiltonian in the doubled Hilbert space. 
Focusing on a concrete example of the one-dimensional transverse-field Ising model as a target Hamiltonian and various types of noise affecting the ITE, we employed DMRG to compute the ground state of the effective Hamiltonian and applied the obtained results to study the SSB order and the associated phase transition in the mixed steady state of the noisy ITE.
In addition, for comparatively small system sizes we directly simulated the evolution of the density matrix under the noisy ITE to show that our conclusions hold in the case of a small but finite Trotter step.

We emphasize that the resilience to noise of the ITE-based ground state phase preparation comes from the presence of the gap in the Liouvillian describing the ITE, or, equivalently, in the effective Hamiltonian of the system in the doubled Hilbert space.
This gap protects the system from small perturbations that are caused by noise.
Therefore, we conjecture that the ITE preparation of a certain phase of matter should be robust to noise as long as the gap in the Liouvillian is present for this phase of matter (as well as the required symmetries, if any, are not broken).
This includes symmetry-protected topological phases (SPTs) and topologically ordered phases.
However, we leave a detailed analysis of the effect of noise on the ITE preparation of SPTs and topologically ordered phases for a future study.
In addition, we would like to emphasize that in this work we have only considered gapped phases of matter.
Investigating preparation of ground state phases and its robustness to noise for gapless systems would be an interesting topic for future analysis.
Here we only note that the gaplessness may alter the convergence time of the ITE, which indeed
depends on the size of the gap.
Another interesting topic for a future study is a comparison between the noisy ITE-based ground state phase preparation and noisy adiabatic/quantum phase estimation protocols.

We conclude by discussing implications of our work for practical simulations of the ITE on noisy quantum devices. 
While in this work we modeled the ITE as a black box, an algorithm implementing the ITE, which is a non-unitary protocol, on a device \cite{qite_chan20,sun2020quantum,QITE_h2,qite_nla,Kamakari2022_qite_open,qite_qft,hejazi2024adiabatic_arxiv,VQITE,Jones2019_vqite,Endo20variational,theory_vqs,AVQITE,smqite,Chen2024_avqite_open,Liu2021_pite,lin2021real,Kosugi2022_PITE,mao2023} requires the introduction of certain nontrivial operations, such as mid-circuit measurements and feedback \cite{mao2023}.
Still, we generally expect our results to hold for a realistic protocol implementing the ITE as long as at each step the (noiseless) protocol well approximates an ITE step, and the state is moved closer to the target ground state.
However, not all protocols implement the ITE in that fashion, and it would be interesting to ``open up" the ITE black box and investigate the robustness of phase preparation against noise in various specific ITE quantum algorithms.
Another potential topic for future studies is to investigate how the Trotter error affects the preparation of the ground state order.
In addition, an interesting direction to explore within this context is the use of noise tailoring techniques like Pauli twirling~\cite{Li_Benjamin-PRX-2017,Chen2022} and probabilistic error cancellation/reduction~\cite{mari2021extendingqp,vandenbergProbabilisticErrorCancellation2023,McDonough-AutomatedPER-2022}.
The possible extension of our results to specific implementations of ITE quantum algorithms in concert with such noise tailoring and error mitigation techniques offers the intriguing prospect of a noise-resilient approach to quantum state preparation.

\begin{acknowledgments}

We thank Peter Orth, Haining Pan, Jedediah Pixley, and Justin Wilson for fruitful discussions. This work was supported by the U.S. Department of Energy, Office of Science, Basic Energy Sciences, Materials Science and Engineering Division, including the grant of computer time at the National Energy Research Scientific Computing Center (NERSC) in Berkeley, California. This part of research was performed at the Ames National Laboratory, which is operated for the U.S. DOE by Iowa State University under Contract No. DE-AC02-07CH11358.

\end{acknowledgments}

\bibliography{ref}

\onecolumngrid

\appendix

\newpage

\section{\label{sec:dmrg_details}Numerical details of DMRG calculations}

\begin{figure*}
	\includegraphics[width = 0.98\columnwidth]{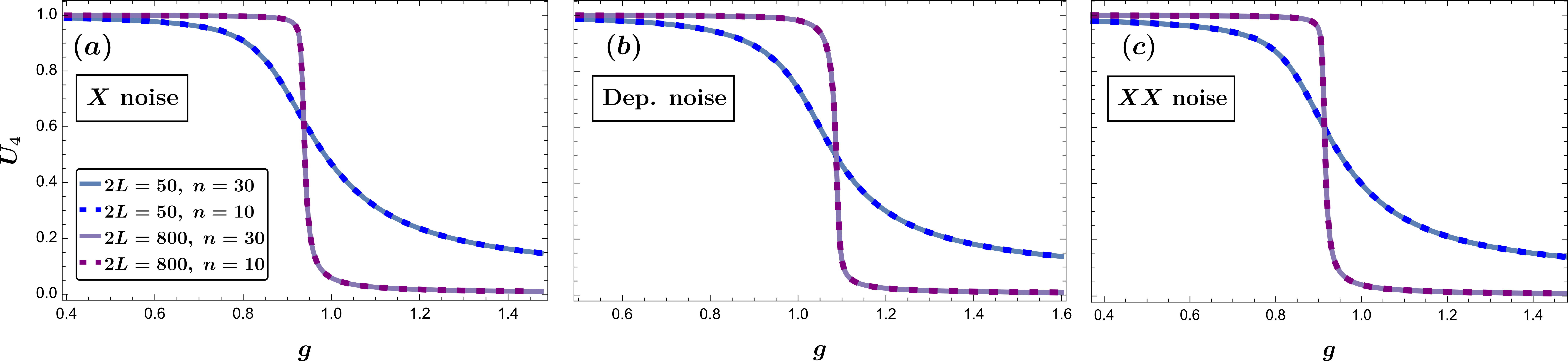}
	\caption{Binder cumulant from DMRG plotted for $n=30$ (solid lines) and $n=10$ (dashed lines) DMRG sweeps, showcasing the convergence of the algorithm. The model and the parameters are fixed to be the same as in Fig.~\ref{fig:dmrg_Pauli_noise}.
}
	\label{fig:dmrg_nsweeps}
\end{figure*}

In this study we utilize \textbf{ITensors.jl} \cite{ITensor_article,ITensor_code} Julia package to perform DMRG calculations.
We refer the reader to its documentation for the precise definition of parameters utilized in our simulations.

To calculate the Binder cumulant \eqref{eq:binder_cumulant} we use the standard DMRG function for open-boundary chains provided by \textbf{ITensors.jl} and compute the ground state of the spin ladder in the matrix-product state (MPS) form.
Starting wavefunction for DMRG is a random MPS with a bond dimension $D_0=30$ and we set the gradual increase of the max bond dimension as $D_{\rm max}=[10,20,100,100,200]$ (unless otherwise specified).
The cutoff in our simulations is set to $c=10^{-10}$.

We perform the DMRG routine for $n=10$ and $n=30$ sweeps.
Figure~\ref{fig:dmrg_nsweeps} illustrates that the results in both cases are the same which showcases the convergence of the algorithm.
The DMRG plots presented in the main text are obtained for $n=30$ sweeps.
Additionally, we checked that improving other DMRG parameters, namely further increasing the bond dimensions $D_0$, $D_{\rm max}$ or reducing the cutoff $c$, does not meaningfully alter our results.

To ensure that within the ordered phase the computed ground state is in the correct sector of the $\mathbb{Z}_2\times \mathbb{Z}_2$ symmetry of the Hamiltonian [which is needed to give a non-zero overlap with the state $|I\rangle\!\rangle$ when calculating $U_4$, see discussion in the paragraph below Eq.~\eqref{eq:H_12_X}], we calculate the expectation value $\langle\!\langle Z_iZ_{i+1}\rangle\!\rangle$ and only accept the computed ground state for which this expectation value is positive.
Here indices $i$ and $i+1$ correspond to the opposing sites on the spin ladder.
If the expectation value is negative, we discard the calculated ground state and redo the DMRG routine for a new random starting MPS. 

Once the ground state of the spin ladder is calculated, we employ Eq.~\eqref{eq:obs_general} with $|I\rangle\!\rangle$ and $O$ written in the MPS and the matrix-product operator (MPO) form, respectively, to compute the expectation values of $m^2$ and $m^4$ needed to obtain the Binder cumulant.

To compute the first excited state (and extract the energy gap $\Delta$), we perform DMRG to minimize the energy while simultaneously minimizing the overlap with the ground state.
Since in the ordered phase the ground state of the spin ladder is four-fold degenerate, we need to obtain the five lowest-energy states to calculate the gap.
When calculating the excited states, we set the DMRG weight parameter to $w=1.0$, which is sufficient since the gap is $\Delta<1.0$ in the region of our interest.

To compute fidelity \eqref{eq:fidelity} we perform DMRG for the Ising chain with and without the noise-induced couplings and compute the respective ground states in the MPS form.
Calculating overlap \eqref{eq:rho_overlap} is then straightforward.
To ensure that both ground states are within the same symmetry sector, we add a small symmetry-breaking perturbation $H_{\rm{SB}}$ at the ends of the spin ladder:
\begin{equation}
    H_{\rm{SB}} = -\lambda_{\rm{SB}}\sum_i Z_i,
\end{equation}
where indices $i$ correspond to the four sites at the ends of the spin ladder.
Throughout our simulations we set $\lambda_{\rm{SB}}=0.1$.
We note that the effect of the perturbation $H_{\rm{SB}}$ disappears in the thermodynamic limit.

\section{\label{sec:gap_CF}Scaling of the gap near critical point and of the correlation function at criticality}

\begin{figure*}
	\includegraphics[width = 0.87\columnwidth]{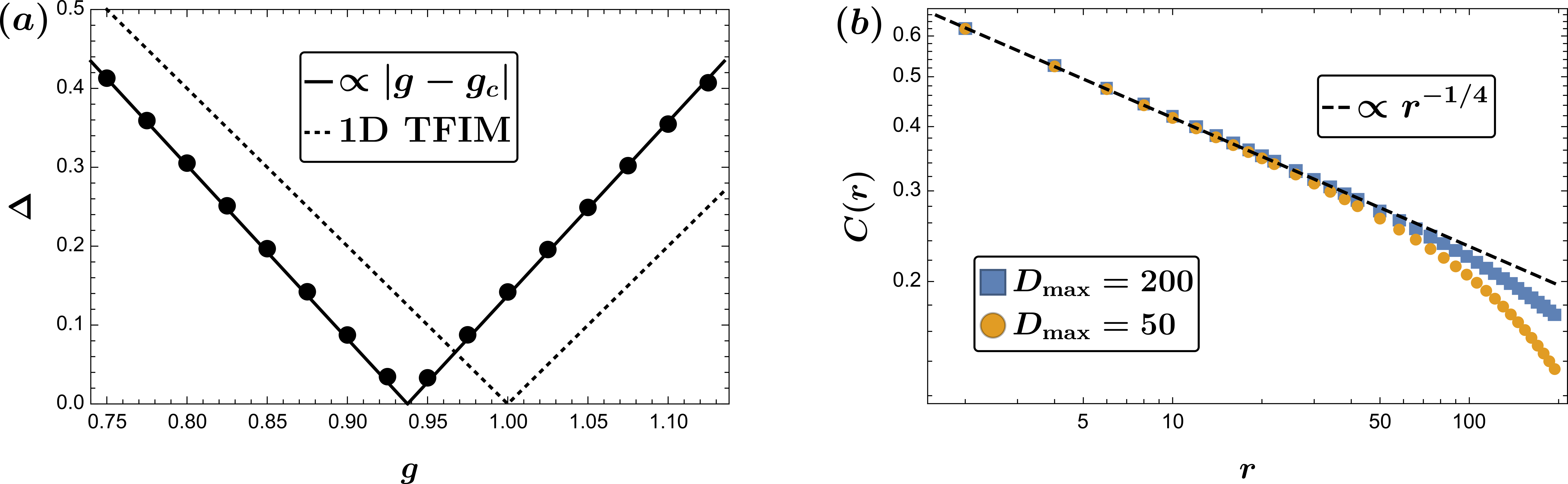}
	\caption{(a) Energy gap near criticality. Solid lines represent the dependence $\Delta\propto |g-g_c|$ with $g_c=0.9375$, while dotted lines display the gap dependence for the (noiseless) 1D transverse-field Ising model, $\Delta\propto |g-1|$. (b) Scaling of the correlation function \eqref{eq:CF_App} with distance at the critical point $g_c=0.937$ for the maximum MPS bond dimensions $D_{\rm max}=200$ and $D_{\rm max}=50$. The correlation function is computed between points situated symmetrically with respect to the center of the spin ladder. Both figures are computed using DMRG for the model with the single-qubit bit-flip noise with $\lambda=0.1$ and system size $2L=800$.}
	\label{fig:dmrg_gap}
\end{figure*}

The energy gap near critical point scales as $\Delta\propto |g-g_c|^{z\nu}$, where $z$ is the dynamical critical exponent and $\nu$ is the correlation length critical exponent.
Figure~\ref{fig:dmrg_gap}(a) illustrates the energy gap as a function of $g$ for the single-qubit bit-flip noise, from which we extract $z\nu=1$.
At the same time, using finite-size scaling of the Binder cumulant (see Section~\ref{sec:Pauli} of the main text) we have obtained $\nu=1$, which together with the results for the gap allows us to extract $z=1$.
This value is consistent with the transition belonging to the 2D Ising universality class.

Furthermore, the connected correlation function,
\begin{equation}
    C(i-j)=\langle\sigma^z_i\sigma^z_j\rangle - \langle\sigma^z_i\rangle\langle\sigma^z_j\rangle,
    \label{eq:CF_App}
\end{equation}
scales as a power-law at critical point:
\begin{equation}
    C(r)\propto r^{-\eta}.
\end{equation}
Figure~\ref{fig:dmrg_gap}(b) demonstrates $C(r)$ calculated using DMRG for the single-qubit bit-flip noise.
Since MPS follow area law of the entanglement entropy, whereas at criticality the entanglement entropy scales logarithmically with system size, it is impossible to capture power-law decay of $C(r)$ over extended distances for a fixed bond dimension (see, for example, Fig.4.12a in Ref.~\cite{Evenbly2013}) -- at a certain $r^{\ast}$ ($r^{\ast}\approx 50$ in Fig.~\ref{fig:dmrg_gap}b) the correlations start to decay exponentially.
Nevertheless, power-law decay over smaller values of $r<r^{\ast}$ allows us to extract the critical exponent $\eta=1/4$, which is consistent with the 2D Ising universality class.
Furthermore, we show that exponential decay sets on at a smaller value of $r^{\ast}$ if we reduce bond dimension of the MPS representing the ground state from $D_{\rm max}=200$ to $D_{\rm max}=50$ (in the latter case the gradual increase of the bond dimension throughout the DMRG routine is set to $D_{\rm max}=[10,20,50]$).
This is expected for critical systems with power-law decaying correlations~\cite{Evenbly2013}.

\section{\label{sec:fidelity_syst_size}Scaling of fidelity with system size}

\begin{figure*}
	\includegraphics[width = 0.9\columnwidth]{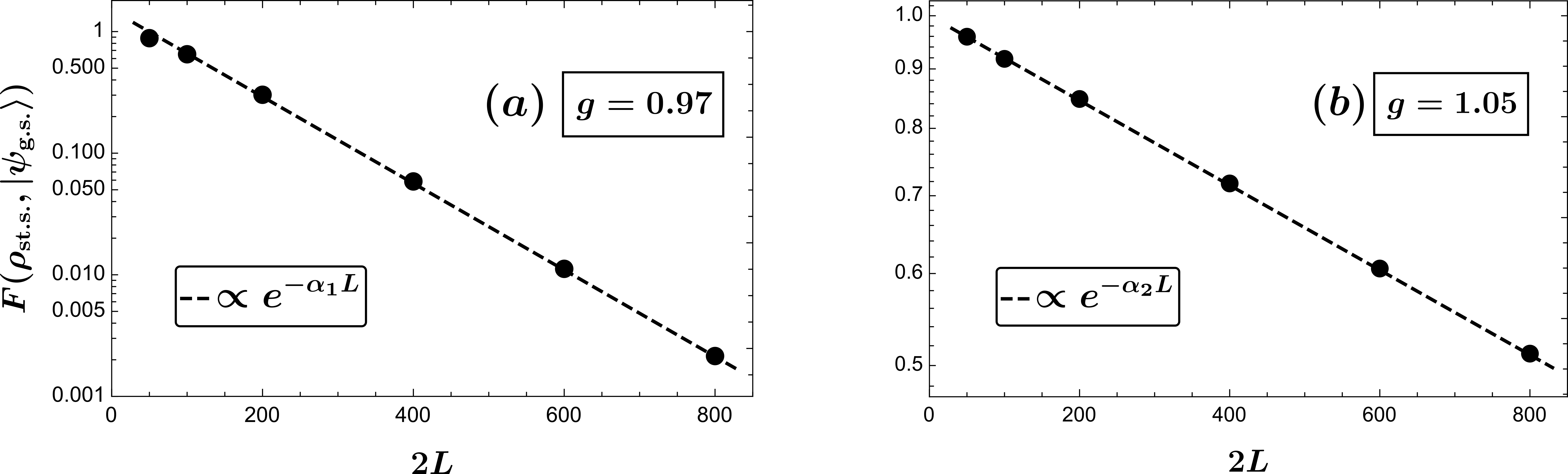}
	\caption{Fidelity between the mixed noisy steady state $\rho_{\rm st.s.}$ and the pure target ground state $|\psi_{\rm g.s.}\rangle$ plotted as a function of system size. The fidelity is computed using DMRG with open boundary conditions via Eqs.~\eqref{eq:fidelity}--\eqref{eq:rho_overlap}. The noise-induced coupling in the effective Hamiltonian is set to $\lambda=0.1$. A small symmetry-breaking field is added at the ends of the spin ladder to ensure that the DMRG algorithm finds a symmetry-broken ground state within the ground-state manifold. The dashed line is a fitted exponential decay with $\alpha_1=0.0082$ and $\alpha_2=0.00084$ for the panels (a) and (b), respectively.}
	\label{fig:fidelity_exp}
\end{figure*}

We plot fidelity as a function of system size in Fig.~\ref{fig:fidelity_exp}.
Figure~\ref{fig:fidelity_exp}(a) demonstrates scaling of the fidelity within the region between the noiseless and the noisy critical points.
Within that region we observe a clear exponential decay of the fidelity to zero with system size.

Scaling of the fidelity with system size outside of the region between the critical points is depicted in Fig.~\ref{fig:fidelity_exp}(b).
While the behavior here is consistent with an exponential decay, the range of available system sizes precludes us from making a definitive conclusion on the scaling.
Still, we expect the fidelity to decay to zero in the thermodynamic limit, $L\to\infty$, due to the orthogonality catastrophe.

\section{\label{sec:fidelity_AD}Fidelity results for the FM Ising chain in the presence of the AD noise}

\begin{figure}
	\includegraphics[width = 0.47\columnwidth]{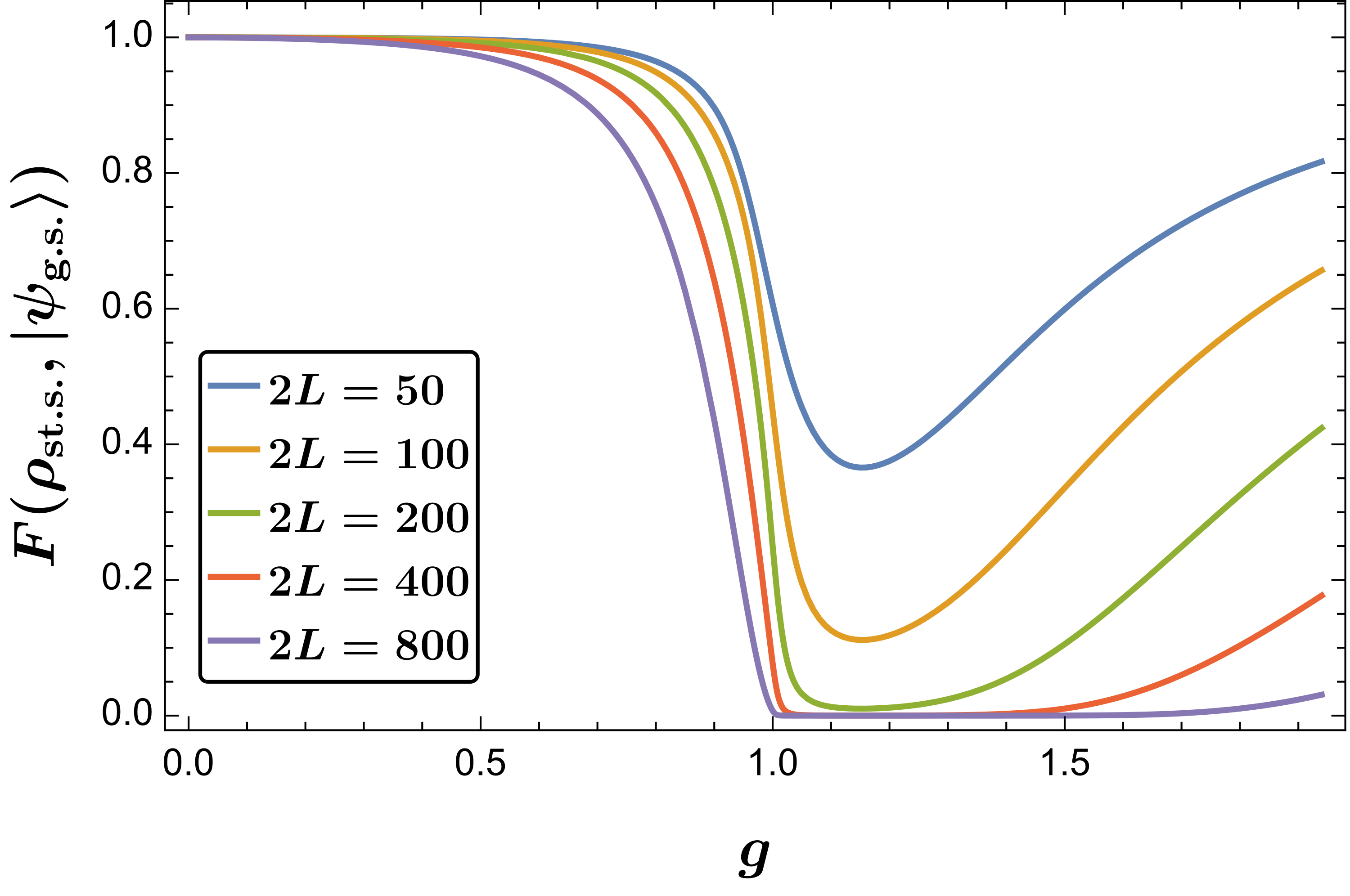}
	\caption{Fidelity between the mixed noisy steady state $\rho_{\rm st.s.}$ and the pure target ground state $|\psi_{\rm g.s.}\rangle$ as a function of the transverse field for the FM Ising chain and the AD noise with $\lambda_z=\lambda_+=0.4$. A small symmetry-breaking field is added at the ends of the spin ladder to ensure that the DMRG algorithm finds a symmetry-broken ground state within the ground-state manifold.}
	\label{fig:fidelity_AD}
\end{figure}

We plot fidelity between the target ground state and the steady state of the noisy ITE for the FM Ising chain in the presence of the AD noise in Fig.~\ref{fig:fidelity_AD}.
While the absence of the SSB order in the noisy case is clearly showcased by the Binder cumulant plots of Fig.~\ref{fig:dmrg_AD}(a), the fidelity plots do not reveal this absence.
Instead, the fidelity is close to one near $g=0$, which is expected due to the fact that the eigenstates of the spin ladder in the noisy and the noiseless cases are the same at $g=0$.
The fidelity then decreases with the increasing $g$, reaching minimum near $g\gtrsim 1$ and then rebounds with the further increase in $g$, rising to $F\approx1$ at $g\to\infty$.
The dip in the fidelity is larger for larger system sizes, similarly to the plots of the fidelity in Fig.~\ref{fig:fidelity} where the noisy system possessed a SSB phase.
However, the dip appears to be wider in the present case.
Note that we expect the fidelity to decay to zero in the thermodynamic limit, $L\to\infty$.
The precise connection between the presence/absence of the SSB order under noise and the state preparation fidelity is an interesting subject for a future study.

\end{document}